\documentclass[%
 reprint,11pt,onecolumn,
 amsmath,amssymb,
 aps,
]{revtex4-1}
\usepackage{comment} % enables the use of multi-line comments (\ifx \fi) 
\usepackage{lipsum} %This package just generates Lorem Ipsum filler text. 
\usepackage{fullpage} % changes the margin
\usepackage{amsmath}
\usepackage{graphicx}% Include figure files
\usepackage{dcolumn}% Align table columns on decimal point
\usepackage{bm}% bold math
\begin{document}

%\preprint{APS/123-QED}

\title{Unrivaled Quantum Vacuum in the Primordial Universe}% Force line breaks with \\

\author{Davood Momeni}
% \altaffiliation[Also at ]{  Center for Space Research, North-West University, Mafikeng, South Africa 
% }
%\\ \&
 %Tomsk State Pedagogical University, TSPU, 634061 Tomsk, Russia.}%Lines break automatically or can be forced with \\
% \email{davoodmomeni78@gmail.com}
\affiliation{% 	
Department of Physics, College of Science, Sultan Qaboos University, P.O. Box 36,
Al-Khodh 123, Muscat, Sultanate of Oman
}%
%\affiliation{ Center for Space Research, North-West University, Mafikeng, South Africa }


%\affiliation{
%\affiliation{
%\affiliation{%


\date{\today}% It is always \today, today,

\begin{abstract}
In a primordial universe pre(post)-inflationary era , there could be  phases of early universe  made of cold gas  baryons, radiation and early post inflationary cosmological constant. I showed that in the baryonic epoch, the quantum vacuum is unique. By using the standard quantization scheme for a massive  minimally coupled scalar field  with maximal conformal symmetry in the classical   spacetime, I demonstrated that the scalar modes had an effective mass $m_{eff}^2\approx 0$ (or $m_{eff}^2\approx constant$). This argument  validated when the conformal time $\eta$ kept so close to the inflation ending time $\eta=\eta_c$. The energy density of the baryonic  matter diverged at the inflation border and vanishes at the late time future. Furthermore I argued that at very early accelerating epoch when the radiation was the dominant part in the close competition with the early time cosmological constant, fine tuned mass of the scalar field $m\propto \sqrt{\Lambda}$ also provided a unique quantum vacuum. The reason is that the  effective mass  eventually is vanished.
A remarkable observation was that all the other possible vacuum states "squeezed" eternally. 
\begin{description}
\item[PACS numbers]
04.60.-m; 04.20.-q.
\end{description}
\end{abstract}

\pacs{Valid PACS appear here}% PACS, the Physics and Astronomy
\maketitle

%\tableofcontents
\section*{Introduction}
Vacuum state in quantum field theory in curved spacetimes doesn't have a unique definition. As we know,in the typical model of scalar field on curved background, once we determine  the pair of creation and annihilation operators  defined as $\{a_k^{-},a_k^{\dagger}\}$ in the momentum space $\vec k$, the possible (and not 
the physical )vacuum state $|0>$ is defined as the eigenstate of  all annihilation operators $a_k^{-}|0>=0,\forall k$. The other alternative approaches to find the vacuum state are the minimization of  instantaneous energy or  minimize the average  for a certain period of time or to  minimize the number of particles with respect to some other vacuum states. 
%%%%%%%%%%55
In addition to the commutation relations between operators $\{a_k^{-},a_k^{\dagger}\}$  we also need to choice an appropriately selection of the mode functions. If we find two sets of annihilation operator  formally as $a_k^{-},b_k^{-}$, we are ending with two copies of the corresponding vacuum namely $|0>_{a},|0>_{b}$. As a result, the particle interpretation is strictly depending on the choice of  mode functions. This statement goes back mainly to the non existence of a plane wave as a general solution for a simple scalar field wave equation in an arbitrary curved spacetime. Only in the Minkowski flat spacetime , the mode frequencies are  time-independent and we can recognize a unique basic function (the plane waves) as quantum vacuum and only in Mikowski spacetime this mode function remains the vacuum mode  at all times i.e, $a_k^{-}|0>_{t=0}=a_k^{-}|0>_{t\neq 0}=0,\forall t\in \mathcal{R}^{+}$. 
Here it is arising an  interesting  question :  does  exist any non flat cosmological spacetime with only one possible vacuum state ?\par 
%%%%%%%%%%%%%%%%%%%%
In this letter I addressed this question by considering a very specific cosmological model in the primordial universe. This  era initiated after the inflation ended. We supposed that this hypothetical phase could be  made of a cold gas of baryons. It was Zel’dovich who showed that the general equation of the state (EoS) for such Universe can be described by the stiff matter era $p=\rho$, where $p$ is the pressure and $\rho$ is the energy density \cite{Zeldovich:1972yb},\cite{Zeldovich:1962yb}. The cosmological epoch  can be considered as an ultra-relativistic regime for a general class of the EoS at $T= 0$ . In the Ref. \cite{Chavanis:2014lra})  a complete analysis of exact cosmological solutions in this family of fluids presented.
 In an earlier paper \cite{Odintsov:2019evb},  the stiff equation the state (EoS) was generated by a light axion field, and the stiff equation of the state  was post inflationary. In the context of the modified theories of gravity as well, a stiff era can also be justified by geometric $f(R)$ gravity terms for example in Ref. \cite{Odintsov:2017cfr}. Furthermore using an axionic field, a string theory motivated axion-like particles (ALP) (not the QCD one ), it was  shown that the axionic field contributes to the early cosmology after inflation ended  with  a stiff fluid EoS at  \cite{Odintsov:2019evb}. However there is a difference between this axionic scenario and our cold gas scenario. The difference is that in the Zel’dovich scenario for stiff fluid , the matter fluid component is a stiff cold baryon gas, whereas in the ALP  case the stiff matter fluid is the axion itself.
  The possibility to have axion as a Bose-Einstein condensate (BEC) proposed firstly in Ref. \cite{dm1}. A key observation was that thermalized cold dark matter axions form a Bose-Einstein condensate. The axion field proposed in Ref. \cite{dm1} was self coupled. There is a possibility to have a stiff pre-inflationary era as it was proved in Ref. \cite{Odintsov:2019evb}.  In Ref. \cite{Odintsov:2019evb} the vauthors showed that the energy density associated to the scalar inflation for a Pre-inflationary era can be approximated as $\rho_{\phi}\sim a^{-6}$ where $a$ is the scale factor for the cosmological background. Following the field equation in the slow roll approximation by neglecting the radiation field and keeping only the scalar field density,they concluded that the scalar field energy density has same order as the axion field i.e, $\rho_{\phi}\sim\rho_{a
 }$. This form for the energy density of the axion field  coincides with the   stiff matter fluid. It is clearly shown that  the model predicts a stiff
matter era. But this stiff fluid era is a pre -inflationary era. Furthermore there is a difference between this pre-inflationary still fluid and the one we considered in this paper as a post inflationary cold fluid. The model we studied is the cold stiff fluid and probably can be explained using the BEC scenario rather than the other auxiliary fields theories. The 
 Zel’dovich stiff phase as we studied in this work appeared in the primordial Universe as we considered here. Furthermore  the Zel’dovich matter fluid is extremely 
cold and it is baryonic , while the stiff fluid proposed in the important Ref. \cite{Odintsov:2019evb} is the axion.  One can show that the axionic stiff fluid contributes to the  late-time era as well as the pre-inflationary primordial era. But for post inflationary epoch the stiff fluid with baryonic origin will be the dominated one.
 Furthermore the classical general relativity can be naturally extended to modified gravity in the form of $f(R)$ form by including the string theory motivated misalignment axion like particles in Ref. \cite{Odintsov:2020nwm}.  The axion dark matter proposed in the above references can be used to probe low-mass particle dark matter. Either the stiff fluid dark matter has been a component of an axion stringy fluid scenario as it was clearly explained in the Refs.  \cite{Odintsov:2019evb} and  \cite{Odintsov:2020nwm} or it came from another scenario different than the BEC scenario, still after the inflation ended, there is no serious problem to consider a tiny cold matter as a baryonic matter content. Technically speaking, the energy density of the baryonic stiff fluid (post inflationary component as we assumed in this work)  is much more smaller than any other energy density of similar fluids (for example axion stiff fluid) in the pre-inflation.    
\par
The aim is to quantize a simple scalar field theory on the classical background arisen from the Zel’dovich's matter content. The  quantization scheme completed in short time interval after the inflation ended. Furthermore, I investigate a mixed phase of early universe, when the universe filled by the radiation and undergoes an acceleration expansion. After taking into account the post inflationary acceleration I consider cosmological constant term.  As we knew, cosmological constant term $\Lambda$ is given as the vacuum energy for the vacuum spacetime. A simple reason to consider $\Lambda$ as a potential candidate for vacuum is the energy-momentum tensor of vacuum expressed in the unique Lorentz invariant form as $T_{\mu\nu}=\Lambda g_{\mu\nu} $ here $g_{\mu\nu}$ defines the Lorentzian metric for curved manifold of the cosmological spacetime. Such early time cosmological constant appeared in the induced gravity scenario in the form of a regulated formulae for $\Lambda$ (see \cite{Visser:2002ew} for modern translation of this idea). The aim is to prove that quantum vacuum has a unique definition for both baryonic and radiation-cosmological constant epochs. 
I use the simple quantization method for massive non minimally couple scalar field on the classical background of the conformally flat metric. The scale factor governs by the Einstein field equation and the scalar field equation of motion simple reduce to the wave equation with effective potential. In the first example I show that cold baryonic matter dominated post inflationary epoch provides a negligible effective mass at cosmological time close to the ending of inflation. Consequently the eigenmodes are reduced to Minkowskian flat modes. In the second example I investigate the effective mass for radiation-$\Lambda$ post inflationary model. It also gives us a tiny effective mass for some values of the scalar field mass. I prove that if the mass of scalar field opted as the fine tuned $\Lambda^{1/2}$, then the effective mass again vanishes. As a result we will end by a unique quantum vacuum.

%The Introduction section, of referenced text\cite{Figueredo:2009dg} expands on the background of the work (some overlap with the Abstract is acceptable). The introduction should not include subheadings.

\section*{Geometry for cold gas baryon phase of the
	early universe }

To have proceed, we have to opt our geometry of the spacetime. We are writing down the set of Friedmann equations with such EoS for a curved, homogeneous and isotropic  Friedmann-Lemaître-Robertson-Walker (FLRW) spacetime characterized by the metric in  conformal time $\eta$ :
\begin{eqnarray}
&& ds^2=a^2(\eta)(d\eta^2-d\vec x^2)\label{metric}
\end{eqnarray}
where $x^\mu\equiv (\eta,\vec x)$. Because our scalar field started after the inflation ended, we suppose that $\eta\geq \eta_c$. We end to the pair of  nonlinear second order differential equations for scale factor and continuity equation,  
\begin{eqnarray}
&&(\frac{a'}{a})^2=\frac{\kappa^2}{3} a^2\rho(\eta) \label{feq1}\\&&
\rho'+\frac{6a'}{a}\rho(\eta)=0
\label{feq2}
\end{eqnarray}

where the prime $'$ denotes derivative with respect to $\eta$ and  $\kappa^2=8\pi G $ is gravitational coupling constant. To compare it with the standard FLRW equation in the cosmological time $t$, we remember that the cosmological Hubble parameter $H=\frac{\dot{a}}{a}$ turns to the $H=\frac{a'}{a^2}=\frac{\mathcal{H}}{a}$ here by $\mathcal{H}$ we mean Hubble parameter in the conformal time $\eta$ defined as $\mathcal{H}=\frac{a'}{a}$.  The $\eta-\eta$ component for Ricci tensor $R_{\eta\eta}=-3(\ln a)''$ and the Ricci scalar $R=-\frac{6}{a^2}((\ln a)^{"}+(\ln a)'^2) $, consequently the $\eta\eta$ component of the Einstein tensor $G_{\eta\eta}=R_{\eta\eta}-\frac{R}{2}g_{\eta\eta}=3(\ln a)'^2$, furthermore the $\eta-\eta$ component for the energy-momentum tensor component $T_{\eta\eta}=g_{\eta\eta}\rho(\eta)$ (see \cite{Iihoshi:2007uz} for a complete study of the geometrical quantities for a conformal cosmological metric). With these expressions we can find the  first Friedmann equation as given in eq.(\ref{feq1}). We notify that the continuity equation doesn't change by passing from cosmological to the conformal time.

%(can be identified as the mass for  next coming scalar field in our scenario).
If we integrate the second differential equation, we obtain \begin{eqnarray}
&& 
\rho(\eta)=\rho_c(\frac{a_c}{a})^6\label{rho}
\end{eqnarray}
here $\rho_c=\rho(a_c)=\rho(\eta=\eta_c) $ denotes the value of density at the inflation ending time. Note that at the early epoch this baryonic matter is the dominant part at least for  short time interval i.e. when $\eta\to \eta+c+\delta\eta$ where $\delta\eta\ll \eta_c$.
By plugging it into the eq. (\ref{feq1}) we can integrate the latter differential  equation.
The general exact solution for the scale factor can be obtained
\begin{eqnarray}
&& a(\eta)=\sqrt{\frac{2\kappa \sqrt{\rho_c}a_c^3}{\sqrt{3}}}|\eta-\eta_c|^{\frac{1}{2}}\label{a}.
\end{eqnarray}
The short early universe covered by $\eta\to\eta_c$ when $a\to a_c$ and late future $a\to \infty$ will be happen at  time  $\eta\to \infty$. Note that the density function
eq. (\ref{rho}) behaves as $\rho(\eta)\propto|\eta-\eta_c|^{-3} $ diverges at early times $\eta\in(\eta_c,\eta_c+\delta \eta]$ will vanish in future. 
If such phase of the early universe existed, contained  the radiation era with $p=\frac{1}{3}\rho$, the dust matter era $p=0$, and the dark energy era $p=-\rho$. A remarkable observation is that the baryonic matter density dominated over all the other  contents.
This type of stiff matter cosmological model is strictly connected to a scenario for early dark matter  in which dark  is composed  of relativistic self-gravitating BECs. Note that in this cosmological model ,the energy density of the stiff matter $\rho$ can be positive or negative. The sign of  energy density is depending on the type of  self-interaction. Furthermore this cosmological era corresponds to a Universe filled 
a perfect fluid at zero temperature $T = 0$ (or some low regimes). Effective thermodynamics model for such Universe  described by a polytropic EoS. If we determine the relation between  energy density and the rest-mass, we observe that the EoS reduces to a stiff EoS  $p=\rho$ describes a cold gas of baryons in Zel’dovich model . A remarkable fact about Zel’dovich model is that  
the velocity of sound is equal to that of light. Consequently the model naturally avoid from any ghost or superluminal propagating mode.
\par
%%%%%%%%%%%%%%%%%%%%%%%%%%%%%%%%%%%%%%
It is illustrative to show that metric with the scale factor  eq. (\ref{a}) provides a positive variable (differ from the  de Sitter ) Ricci scalar given by $R\propto (\eta-\eta_c)^{-3}$ and non vanishing components of the Riemann tensor $R^{\eta}_{\eta b b }\propto R^{c}_{c b b }\propto(\eta-\eta_c)^{-3} , b,c=1,2,3 $. The metric given in  eq. (\ref{metric}) is  non flat and distinct from the de Sitter exact cosmological model. This metric provides our classical background where the quantum scalar field propagates. To investigate dynamics of the scalar field on this classical background we follow the method proposed in \cite{Mukhanov}. 

\section*{ Quantum Scalar field on classical homogeneous and isotropic backgrounds}
Let us start considering a minimally coupled scalar field $\phi(x^{\mu})$ in a curved cosmological background presented in eq. (\ref{metric}) with scale factor eq. (\ref{a}),
the action is
\begin{eqnarray}
&&S=\int \sqrt{-g}d^4x \Big(\frac{1}{2}\phi_{;\mu}\phi^{;\mu}-\frac{1}{2}m^2\phi^2
\Big) \label{action}.
\end{eqnarray}
here $\phi_{;\mu}=\partial_{\mu}\phi=\nabla_{\mu}\phi$ and $m$ is mass of the scalar field (or scaleron). The field equation using an auxiliary field $\chi=a(\eta)\phi$ reduces to the equation of motion for auxiliary field $\chi$ in the Minkowski spacetime for a time dependent effective mass 
\begin{eqnarray}
&&m^2_{eff}=m^2a^2-\frac{a''}{a}
\label{meff}.
\end{eqnarray}
By plugging the scale factor given in eq. (\ref{a}) in the above equation  we can show that for $m=0$ we can vanish the effective mass 
at time $\eta=\eta_c$ . Since the baryons produced after the inflation ended, the massless scalar field in our scenario lives on the border when  time scale of inflation is identified as $\eta_c$. As a result, scalar field theory with a massless scaleron reduces to the massless scalar field theory on a flat space. Note that although our background metric is non flat but the reduced action still is  defined on Minkowski spacetime and with vanishing effective mass gives modes with uniform frequency and independent from the time $\eta$ . There is another possibility to vanish the effective mass for all times close to $\eta-\eta_c\propto m^{-3/4}$ but we are interesting to have a scale invariant scenario with no specified time scale , we prefer a massive scalar field instead of finite flatness condition where the effective mass vanish at  conformal time scales.\par

\par
One can set $m_{eff}\approx0$ and then 
by expanding the auxiliary field $\chi$ in Fourier modes
\begin{eqnarray}
&&\chi(\vec x,\eta)=\int\frac{d^3k}{(2\pi)^{3/2}}\chi_k(\eta)e^{i\vec k\cdot \vec x}
\label{chik}.
\end{eqnarray}
the decoupled equations of motion for the mode $\chi_k(\eta)$,
\begin{eqnarray}
&&\chi''_k+k^2\chi_k=0
\label{chik}.
\end{eqnarray}
here $k=|\vec k|$(with respect to the Euclidean norm not the curved one). This is  a harmonic oscillator equation. All modes $\chi_k(\eta)$ with $k=|\vec k|$ are complex exponential solutions. If we choice a normalized  mode function $v_k(\eta)=\frac{e^{ik\eta}}{k}$ subjected to the normalization condition $\Im (v'_kv_k^{*})=1$(by $^*$ we mean the complex conjugate ), we can show that the exact solution for $\chi_k(\eta)$ can be expressed as
\begin{eqnarray}
&&\chi_k(\eta)=\frac{1}{\sqrt{2}}\Big(a_k^{-}v_k^{*}(\eta)+a_{-k}^{\dagger}v_k(\eta)
\Big)
\label{chik}.
\end{eqnarray}
Note that since $\chi_k$ is a real function, $\chi_k^{*}=\chi_{-k}$ and $a_{k}^{\dagger}=(a_k^{-})^{*}$. A complete form of the solution for $\chi(\vec x,\eta)$ can be represented as follow

\begin{eqnarray}
&&\chi_k(\vec x,\eta)=\frac{1}{\sqrt{2}}\int \frac{d^3k}{(2\pi)^{3/2}}\frac{1}{|k|}\Big(a^{\dagger}_{k}e^{i(k\eta-\vec k\cdot\vec x)}+a_{k}^{-}e^{-i(k\eta-\vec k\cdot\vec x)}
\Big)
\end{eqnarray}

To make our scenario fully quantized we apply the real-time Heisenberg commutation relations between the field $\hat{\chi}$ and its conjugate momentum $\hat{\pi}$
\begin{eqnarray}
&&[\hat{\chi}(\vec x,\eta),\hat{\pi}(\vec x,\eta)]=i\delta(\vec x-\vec y)
\end{eqnarray}
here $\hat{\pi}=\frac{d\hat{\chi}}{d\eta}=\hat{\chi}'$(read it in the Heisenberg picture). An alternative method is to quantize the field $\chi$ via mode expansion:
\begin{eqnarray}
&&[\hat{a}_k^{-},\hat{a}_{k'}^{\dagger}]=\delta(k-k')\\
&&[\hat{a}_k^{\pm},\hat{a}_{k'}^{\pm}]=0.
\end{eqnarray}
Because the complex field $\chi$ is a set of two real fields, we can also represent the field $\chi$ in the form 

\begin{eqnarray}
&&\chi_k(\vec x,\eta)=\frac{1}{\sqrt{2}}\int \frac{d^3k}{(2\pi)^{3/2}}\frac{1}{|k|}\Big(a^{-}_{k}e^{-i(k\eta-\vec k\cdot\vec x)}+a_{k}^{\dagger}e^{i(k\eta-\vec k\cdot\vec x)}
\Big)
\end{eqnarray}

Now I will show that there is no other vacuum state for our scalar theory in this special spacetime. If we can find two sets of isotropic mode functions $u_k(\eta),v_k(\eta)$, because $u_k(\eta),u_k^{*}(\eta)$ are a basis , the functions $v_k(\eta)$ can be expanded in this new basis ,
\begin{eqnarray}
&&v_k^{*}(\eta)=a_ku_k^{*}(\eta)+\beta_k u_k(\eta)
\end{eqnarray}
here $\alpha_k,\beta_k$ named as Bogolyubov coefficients.  It is impossible to find more than one basis function $v_k(\eta)$ for mode equation eq. (\ref{chik}) 
to be normalized over the entire interval $\eta_c\leq \eta< \infty$. For example if we choice $u_k(\eta)=B\sin(k\eta)$ , the normalization condition can not be satisfied. We conclude that there isn't more than one vacuum (one basis function) for our scalar theory (all the Bogolyubov coefficients vanish ). We mention here that there will be more than a vacuum state if we let time coordinate $\eta\gg\eta_c$ for example if $\eta\to N\eta_c,N\gg 1$. A possible explanation is to divide time interval 
\begin{eqnarray}
&&I=(\eta_c,\infty)=(\eta_c,\eta_c+\delta\eta]\lim_{N\to\infty}\cup_{i=1}^N(\eta_c+\delta\eta,i\eta_c) 
\end{eqnarray}

In the first time interval we have a unique vacuum because the mode functions are only a set. In the remaining partition of the interval it is possible to find more than one set of basis functions in any sub interval given by $I_i=(\eta_c+\delta\eta,i\eta_c)$ subjected to  the normalization condition. But still one can show that the total number of the particles will diverge at limit $\lim_{N\to\infty}$.Now the question is why the other vacuum states are absent?. One can answer it by mentioning that the cosmological  background is filled by a baryonic gas with extra cold regime $T=0$. Except the one vacuum state , the others squeezed by this extreme cold regime. We conclude that there is a unique choice of vacuum for a scalar field theory propagated in a non flat the primordial cosmological background   made of a cold gas of baryons. Because such cold gaseous system can be  made of relativistic self-gravitating Bose-Einstein condensates (BECs) and early dark matter may be composed by  such self-gravitating  short-range interactive BEC systems \cite{dm1}, we should be able to detect this unique vacuum state in an experimental set up easily.\par
%%%%%%%%%%%%%%%
%Now I compute the particle creation rate at times longer than $\eta_c$. For massless scalar field in the Zeldovich's epoch with scale factor  given in eq. (\ref{a}) , the normalized mode functions $\chi_k(\eta)$ admits solutions under the form of Hankel’s functions:
%\begin{eqnarray}
%&&\chi_k(\eta)=\frac{\sqrt{\pi\eta}}{2}\big(\epsilon_1H^{(1)}_0(k|\eta-\eta_c|)+\epsilon_2H^{(2)}_0(k|\eta-\eta_c|)\big)\ \ \epsilon_{1,2}=0,1.
%\end{eqnarray}
%The set of isotropic mode functions for $\epsilon_2=0$ is 
%\begin{eqnarray}
% &&v_k^{*}(\eta)=\frac{\sqrt{\pi\eta}}{4}e^{i\theta_k}\big(H^{(1)}_0(k|\%eta-\eta_c|)-iH^{(1)}_{-1}(k|\eta-\eta_c|)\big).\ \ 
%5\end{eqnarray}
%hence the expression gives  rate of particle production is $|v_k(\eta)|^2$. It shows that particles produced at the Zeldovich's epoch after inflation ending $\eta=\eta_c$ is compatible with structure formation(see \cite{Batista:2007gd} for case of an accelerating Universe).

%\subsection*{Subsection}

%Example text under a subsection. Bulleted lists may be used where appropriate, e.g.

%\begin{itemize}
%\item First item
%\item Second item
%\end{itemize}

%\subsubsection*{Third-level section}

%Topical subheadings are allowed.
\section*{Post inflationary non baryonic  unique vacuum}
In previous section we showed that the cold early baryonic phase leads to a unique Minkowskian quantum vacuum. Another remarkable example of early unique vacuum state is a post inflationary epoch when the universe filled with the early radiation fields with energy density $\rho\propto a^{-4}$ along an early post inflationary accelerating expansion. A possible total energy density for this mixed early phase is given by
\begin{eqnarray}
&&\rho=\frac{3\mu^2}{2\kappa^2}+\frac{3C}{\kappa^2 a^4}.
\end{eqnarray}
 There is an induced radiation term here . It dominated  at the tiny  scales $\rho\propto a^{-4}$. One reason to have radiation term  is that the radiation component  created  in standard scenarios with inflation  after inflation ended. Although we didn't study the role of the radiation component extensively here but we point out that the effect of the radiation to the quantum fluctuations and the origin of particles investigated in the existed literature.  Since we are studying post inflationary epoch, the cosmological constant term is the dominant part because it is needed to keep the Universe accelerating smoothly. Furthermore  the radiation density goes down when the post inflation epoch started and that is because the expansion of the Universe continued. Furthermore we reminder that the radiation components also existed in the rehearing epoch.
It is easy to show that the above energy density corresponds to the FLRW model with radiation field and an effective early cosmological constant $\Lambda\propto \mu^2$ . Furthermore the effective mass given in eq. (\ref{meff}) could be vanish if the free parameter $\mu=m\propto \sqrt{\Lambda}$. A rough estimation for $\mu$ is given as   $\mu\approx $ in natural units. Such scalar field can generate a post inflationary accelerating epoch with an effective cosmological constant $\Lambda$. Since $\Lambda$ is considered as a fine tuned constant, the mass for the scalar field also is fine tuned. A remarkable result is that post inflationary accelerated epoch could be described by a toy scalar field with a unique quantum vacuum. In addition to the former post inflationary cold bayonic phase we have also an additional radiation-accelerating phase with unique quantum vacuum. The question about which vacuum state survives from reheating and other phases after ending inflation can not be answered easily without knowing more about the quantum vacuum  and calculating cosmological parameters. To have $m_{eff}=constant$ instead of making it vanish, we demand the scale factor be  adjusted along the scalar mass $m$ and effective $m_{eff}$. For this purpose we consider the case where we are looking for such a cosmological background which vanishes our effective mass $m_{eff}$. We rewrite the constant  mass expression as follows, 
\begin{eqnarray}
&&a''=-\frac{d}{da}V_{eff}(a),\\&&
V_{eff}(a)\equiv \frac{1}{2}\Big(m_{eff}^2a^2-\frac{m^2}{2}a^4)
\end{eqnarray}
note that in the above effective classical representation, the potential posses three equilibrium points $a_e^{\pm,0}=\pm|\frac{m_{eff}}{m}|,0$, the roots $a^{\pm}$ correspond to unstable vacuum states and only the origin gives  stability  with quantum  energy given by $E_{0}\approx \frac{m_{eff}}{\sqrt{2}}$
%(see the FIG.1). 
We conclude from the above figure  that the inclusion of the scalaron mass $m^2$ term  reduces  the initial energy density $\rho_0$ in eq. (\ref{feq1}).

% \begin{figure}
%\includegraphics[scale=0.3]{fig1.eps}
% \caption{\label{fig1} Effective potential $V_{eff}(a)$ with $m=m_{eff}=1$
%	. The effective potential
%	shows a twin doubly unstable vacuum and a stable vacuum 
%	for $a_e^{\pm,0}=\pm|\frac{m_{eff}}{m}|,0$ respectively. }\label{fig1}
% \end{figure}
As a result one can develop a perturbative scale factor $a(\eta)$ up to the first order,
\begin{eqnarray}
&&a(\eta)\approx A\cos(m_{eff}\eta+\phi_0)+B|\frac{m}{m_{eff}}|\cos(\sqrt{2}m_{eff}\eta+\phi_1)\label{a1}
\end{eqnarray}
here $\phi_{0,1}$ are initial phases, $B=\mathcal{O}(A)$ is a constant  and for sake of the perturbative regime we assume that $|\frac{m_{eff}}{m}|\gg 1$, it implies that the scalar field is super light and loss the mass during epoch.  We mention here that there is no radiation component in the eq. (\ref{a1}). This expression presents a perturbation solution to the effective potential for the scale factor. This scenario is the opposite side of the previous  where we adopt a non flat but still vacuum classical background with a super massive scalaron. This cosmological scenario with bouncing the factor given in eq. (\ref{a1}) reviewed in \cite{bounce}. A remarkable observation is that our proposed scale factor leads to a non vanishing Ricci scalar $R\propto a^{-2}(m_{eff}^2-m^2a^2)$. At the vicinity of  unstable vacuum state we obtain  $R\approx 0$ and spacetime still remains close to flat and to the stable vacuum epoch, when $a\to a_e\to 0$, the space time suffers from a initial singularity $R\to \infty$ as we expected from the big bang scenario as well  an infinite amount for the initial energy density $\rho_c$ in eq. (\ref{feq1}) as it was expected from the primordial model with Zeldovich's EoS presented in this letter.  
%%%%%%%%%%%%%%%%%%%%%%%%%%%%%%%%%%%%%%%%
The  quantum vacuum fluctuations of the  single scalar degree of freedom around the stable critical point $a_e=0$ are compatible with inflationary scenario as a non singular bounce model \cite{inflation}. 
\section{Discussion}
In summary I investigated the problem of uniqueness  the vacuum state in curved cosmological background. There is  ambiguity to define vacuum state in curved spacetime. It refers to the  usual definitions of the vacuum and of “particles with momentum $k$” in the Minkowski spacetime. In flat manifolds the mode decomposition of the scalar field equation of the motion is based on the  plane waves
solutions of the ordinary quantum mechanics. According to the Heisenberg's uncertainty principle, the wave function for a free particle with  momentum $p$ is  a
wave packet with standard variance $ \Delta p$. The  spread of the wave packet should be
sufficiently small $\Delta p\ll p$. This condition is necessary to define an observable   momentum of the particle. According to the uncertainty principle, the spatial size $l$ of the wave packet satisfies  $l\Delta p\sim 1$,as a result we have $l\gg p^{-1}$. When we try to quantize a scalar field on a curved background (here a cosmological metric) , there is a risk to have a significant variation of the geometry (for example the scalar Ricci) across
a region of size $l$. Even if the chance be very low still we can conclude that  the plane waves are becoming  a poor approximation as solutions for the scalar field on such curved backgrounds. Furthermore, we understand that 
the particles with momentum $p$ cannot be defined similar to the Minkowskian case . The notion of a free particle with momentum $p$ is failed. The only possible meaningful way to keep the concept of the free particle is to preserve the metric of  curved spacetime very close to the Minkowski  at the scales $p^{-1}$. In this letter we demonstrated that there is a curved background which is different from the Minkowski spacetime and still it has plane wave as a vacuum state. The only constraint is that the mass of the scalar field should be matched to the free parameter (probably some type of the charge or mass )of the cosmological background. Because the geometry corresponds to a very early cold era of the Universe, our result suggested that the other vacuum states squeezed smoothly during this cosmological evolutionary epoch. We conclude that although the  “vacuum” and “particles” are approximate concepts and for a generic curve background   are  ambiguous , in the primordial epoch of the universe , all  observers are agree on one  type of the  vacuum.
\section*{Acknowledgements}

This work supported by the Internal Grant  has been assigned a code number (IG/SCI/PHYS/20/07)  provided by Sultan Qaboos University .


\end{document}